\documentclass[preprint,noshowpacs,amsfonts,amssymb,amsmath,superscriptaddress]{revtex4}

	\usepackage{bm}

	\usepackage{hyperref}		
	\usepackage[figure]{hypcap}		
	\hypersetup{
	   	bookmarks		= false,		
		unicode		= false,		
		pdfborder		= {0 0 0}		
		pdftoolbar		= true,		
		pdfmenubar		= true,		
		pdffitwindow		= false,     		
		pdfstartview		= {FitH},    		
		pdfnewwindow	= true,      		
		colorlinks		= true,		
		linkcolor		= blue,		
		citecolor		= blue,		
		filecolor		= blue,		
		urlcolor		= blue,		
		linktocpage		= true,		
		pdftitle		= {Electronic band structure of rhenium dichalcogenides},
		pdfauthor		= {SURANI M. GUNASEKERA, DANIEL WOLVERSON, LEWIS S. HART and MARCIN MUCHA-KRUCZY\'{N}SKI},
		pdfsubject		= {},
		pdfcreator		= {},
		pdfproducer		= {},
		pdfkeywords		= {: ReSe\textsubscript{2}, ReS\textsubscript{2}, rhenium dichalcogenides, ARPES, band structure},
}

 	\usepackage{graphicx}

\begin{document}

\title{Electronic band structure of rhenium dichalcogenides}
\author{S.~M.~Gunasekera}
\author{D.~Wolverson}
\author{L.~S.~Hart}
\author{M.~Mucha-Kruczy\'{n}ski}
\affiliation{Centre for Nanoscience and Nanotechnology, Department of Physics, University of Bath, Bath BA2 7AY, United Kingdom}

\begin{abstract}
The band structures of bulk transition metal dichalcogenides ReS\textsubscript{2} and ReSe\textsubscript{2} are presented, showing the complicated nature of the interband transitions in these materials, with several close-lying band gaps. Three-dimensional plots of constant energy surfaces in the Brillouin zone at energies near the band extrema are used to show that the valence band maximum and conduction band minimum may not be located at special high symmetry points. We find that both materials are indirect gap materials and that one must be careful to consider the whole Brillouin zone volume in addressing this question. 
\end{abstract}

\pacs{ignore}

\maketitle

\section*{INTRODUCTION}
\noindent The field of two-dimensional few-layer and monolayer materials has generated intense and sustained interest since the isolation of graphene in 2004 [1], though bulk (three-dimensional) van der Waals layered materials and, in particular, the transition metal dichalcogenides (TMDs), were known and studied much earlier [2]. In the last few years, many more two-dimensional materials have been identified or proposed and there are currently around 1000 candidates for two-dimensional metals, semiconductors, superconductors and charge density wave materials. Besides graphene, the TMDs are still the most actively studied members of this family and many prototype devices (e.g, field effect transistors, sensors, and photo-detectors) based on MoS\textsubscript{2} have been demonstrated successfully, as well as device paradigms being proposed (e.g, spin- and valleytronics) based on the band structure of mono- or few-layer MoS\textsubscript{2} [3-4].\newline

\noindent Although MoS\textsubscript{2} is arguably the archetypal semiconducting TMD, there are many TMDs that offer contrasting properties to MoS\textsubscript{2} and therefore add significantly to the diversity of TMD devices and heterostructures that can be explored [2]. The rhenium chalcogenides ReS\textsubscript{2} and ReSe\textsubscript{2} are prime examples of this [5] since they differ markedly from MoS\textsubscript{2} in crystal structure and symmetry [6], electronic band structure [7], and lattice dynamics [8]. The crystal structure of these compounds is shown in Figure 1, based on early crystal structure determinations \textit{via} X-ray diffraction [6] and optimisation of the structures via first-principles calculations as described in the Methods section. In the ReX\textsubscript{2} structure (X=S, Se) Re atoms group into diamond-shaped clusters of 4 atoms and each cluster is linked to the next one by Re-Re bonds to form chains running along the crystallographic axis direction we define as \textit{a} [9]. The resulting structure has only inversion symmetry and is highly anisotropic in all physical properties in-plane, in contrast to the hexagonal TMDs. The stacking of layers is such that the \textit{c} axis is not perpendicular to the layer plane but \textit{a} and \textit{b} do lie in the layer plane, so that the reciprocal space \textit{\textit{c*}} axis is normal to the real space layers and thus also normal to the principal surface of bulk crystals; \textit{a*} and \textit{b*} lie out of the layer plane. Thus, angle-resolved photoemission experiments (ARPES), which preserve information about electron momentum normal to the sample surface, will probe the valence band dispersion in an approximately planar section through the Brillouin zone that does not contain any of the reciprocal lattice vectors (this is well illustrated in Ref. [10], Figure 2).  \newline

\begin{figure}[t]
\centering
\includegraphics[width=0.5\textwidth]{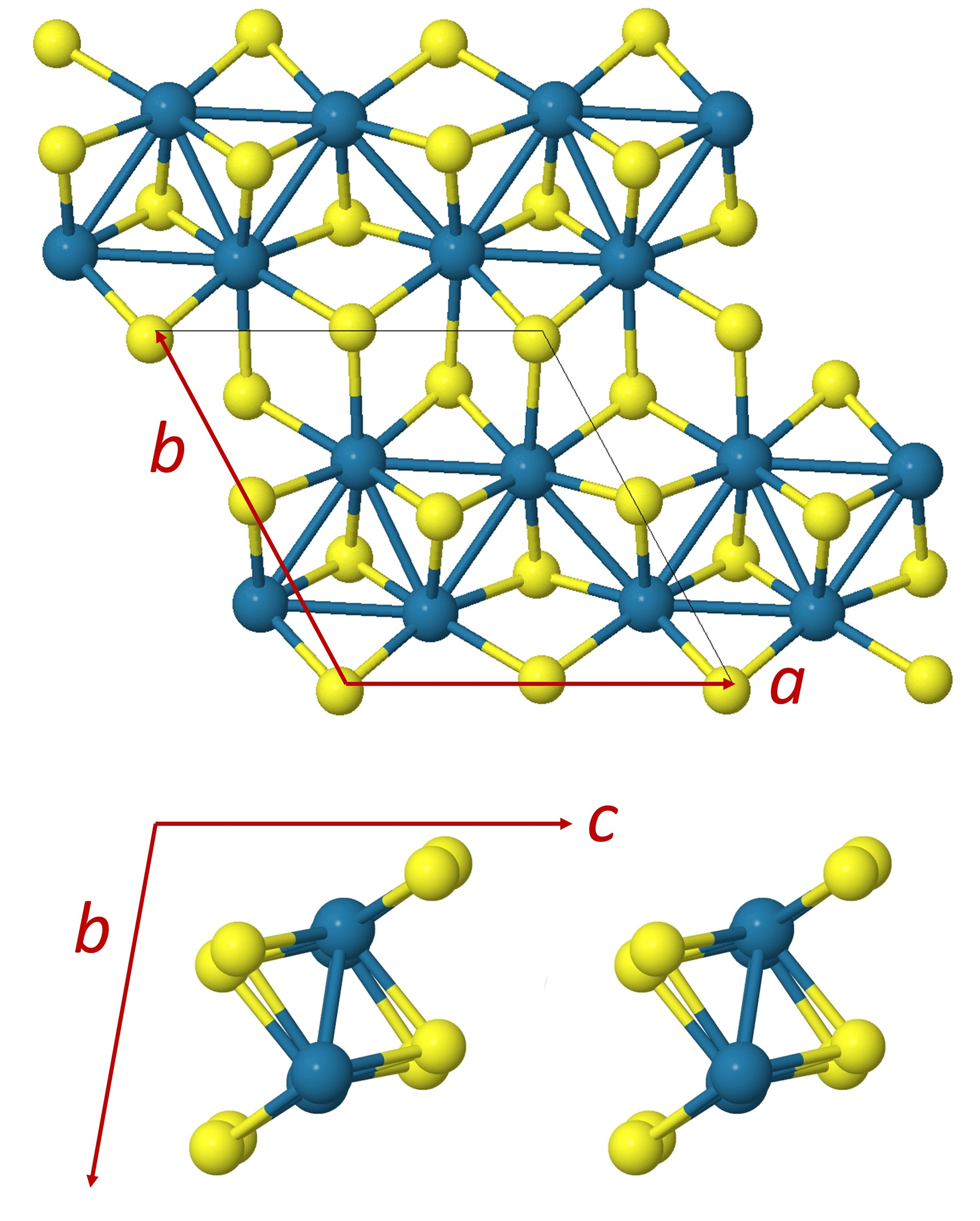}
\caption{Top and side views of the ReX\textsubscript{2} structure with rhenium (sulphur or selenium) atoms shown in blue (yellow); the rhenium atoms lie in the layer plane. The \textit{a} and \textit{b} crystallographic axes are in the layer plane and are defined as shown in the upper panel; the side view (lower panel) is drawn looking along \textit{a} and shows that the \textit{c} axis is not normal to the layer plane. }
\label{fig:figure1}
\end{figure}

\noindent Recently, we have presented experimental data on the electronic valence band structures of ReS\textsubscript{2} [11] and ReSe\textsubscript{2} [12] as determined by nanoscale ARPES, and closely related work has been reported also by other groups, on both bulk ReS\textsubscript{2} [13] and also few-layer ReS\textsubscript{2} [10]. This work addresses controversies over (i) the nature of the bandgaps in the ReX\textsubscript{2} family and (ii) the degree to which the ReX\textsubscript{2} bulk materials can be considered as non-interacting stacked layers. It has become clear that the details of a particular first-principles calculation influence the conclusions about the existence of a direct gap and the locations of the conduction and valence band extrema; these materials present a very flat band structure with a large number of bands in a small energy range (due to the large 12-atom unit cell) and, as we shall show, the band extrema are not necessarily located at any high symmetry points of the Brillouin zone (BZ). Calculations typically focus on high symmetry paths in the BZ and may miss the true band edges; here, we take a different approach, calculating the band energies over the whole volume of the BZ and tracing out constant energy surfaces rather than dispersions. In particular, we present data for the conduction band of bulk ReSe\textsubscript{2}, which has not previously been discussed in detail. 

\section*{COMPUTATIONAL METHODS}
\noindent Electronic band structures were calculated via plane-wave, pseudopotential methods within density functional theory (DFT) using the QUANTUM ESPRESSO package [14]. Structures were derived from published X-ray diffraction (XRD) crystallographic data [6] and were relaxed with respect to both unit cell dimensions and atomic coordinates to give atomic forces of less than 6.1$\times$10\textsuperscript{-3} eV \AA \textsuperscript{-1} (ReS\textsubscript{2}) or 3.5$\times$×10\textsuperscript{-2} eV Å \AA \textsuperscript{-1} (ReSe\textsubscript{2}). Fully relativistic pseudopotentials were used with the projector augmented wave (PAW)  method [15]; pseudopotentials and PAW datasets were constructed using the PSLibrary [16] for the local density approximation (LDA) Perdew-Zunger [17] and generalized gradient approximation (GGA) Perdew-Burke-Ernzerhof [18] exchange-correlation functionals. The valence of Re was taken as 15 (configuration 5s\textsuperscript{2} 5p\textsuperscript{6} 5d\textsuperscript{5} 6s\textsuperscript{2}). Monkhorst-Pack [19] \textit{k}-point meshes of 10$\times$10$\times$10 (ReS\textsubscript{2}) or  8$\times$8$\times$8 (ReSe\textsubscript{2}) were used with kinetic energy cutoffs of, typically, 60 Ry (816 eV); convergence with respect to both of these was checked. In this work we present results using fully relativistic LDA and GGA functionals; we have explored the differences between the results using the LDA and GGA levels of approximation further elsewhere [12, 20]. 

\section*{RESULTS AND DISCUSSION}
\noindent No direct comparison of the band structures of ReSe\textsubscript{2} and ReS\textsubscript{2} over the whole Brillouin zone (BZ) has yet been presented and so we provide such a comparison here. This is useful as a guide to the electronic properties of these materials but also gives insight into the difficulties in making first-principles calculations of band structure in this system; our results show that it is necessary to consider the whole BZ and not to focus only on high-symmetry paths. To make this comparison, we use consistently two types of pseudopotential; although more sophisticated calculations of band structure can be performed, and the exact positions of the band extrema depend on the choice of pseudopotential, the above comments will apply at any level of approximation.

\subsection*{Rhenium sulphide}
\noindent We consider first ReS\textsubscript{2}, being the more well-studied of the two materials, and the material with the wider bandgap. It should be noted that here we take the bulk unit cell to contain just one monolayer and 4 formula units; this contradicts the conclusions of some early XRD studies of ReS\textsubscript{2} [6] but is consistent with more recent XRD [21] and photon-energy dependent ARPES studies [10, 13, 20]; the latter technique probes the valence band dispersion in the direction normal to the sample surface (which is here the layer plane) and is thus sensitive to the lattice periodicity in that direction.\newline

\noindent Figure 2 shows constant energy surfaces for two energies chosen to show (a,c) the conduction (CB) and (b,d) valence band (VB) structures clearly within the full three-dimensional BZ. In both cases, the smaller (red) surface encloses the band extremum and the larger (yellow) surface shows how the bands develop at energies further from the band extremum. To obtain these figures, the band energies were calculated for a grid of points covering the whole BZ. This process is not usually required in more high-symmetry materials and can be computationally expensive; each of our datasets required 24 hours with 64 processors, which is acceptable, but there is a cubic scaling with \textit{k}-space resolution which makes the use of a very fine grid impractical. Therefore, we do not attempt to extract band dispersions from the same dataset, and we are only able to locate the band extrema to the resolution of our \textit{k}-point grid. However, we propose that presenting the band structure in this way is useful, since this process searches for the band extrema without prejudging where they are located. As Fig. 2 makes clear, this calculation implies that both the VBM and CBM are not located at $\Gamma$ or at any particular high-symmetry point on the BZ boundary. \newline

\noindent Most reported calculations of the band structure so far have presented the dispersion only along a special set of directions. In some cases, these have been chosen connecting the $\bar{\Gamma}$, $\bar{K}$ and $\bar{M}$ points of a hypothetical two-dimensional BZ; however, $\bar{K}$ and $\bar{M}$ are not special points in the three-dimensional BZ, but are projections of the latter onto the layer plane [7, 22-25] (and so cannot be represented on Figure 2). In other cases, paths have been chosen in the 3D BZ connecting three-dimensional special points but not exploring the whole BZ [26-27]. The 2D projection onto the layer plane is very useful when considering how the band structure evolves from bulk to monolayer materials and is also necessary when modelling ARPES experiments, in which a 2D section through the 3D BZ is measured. However, by discarding the \textit{\textit{c*}} component of the wavevector, it is impossible to determine whether the band extrema occur at the same crystal momentum values.  \newline

\begin{figure}[t]
\centering
\includegraphics[width=0.7\textwidth]{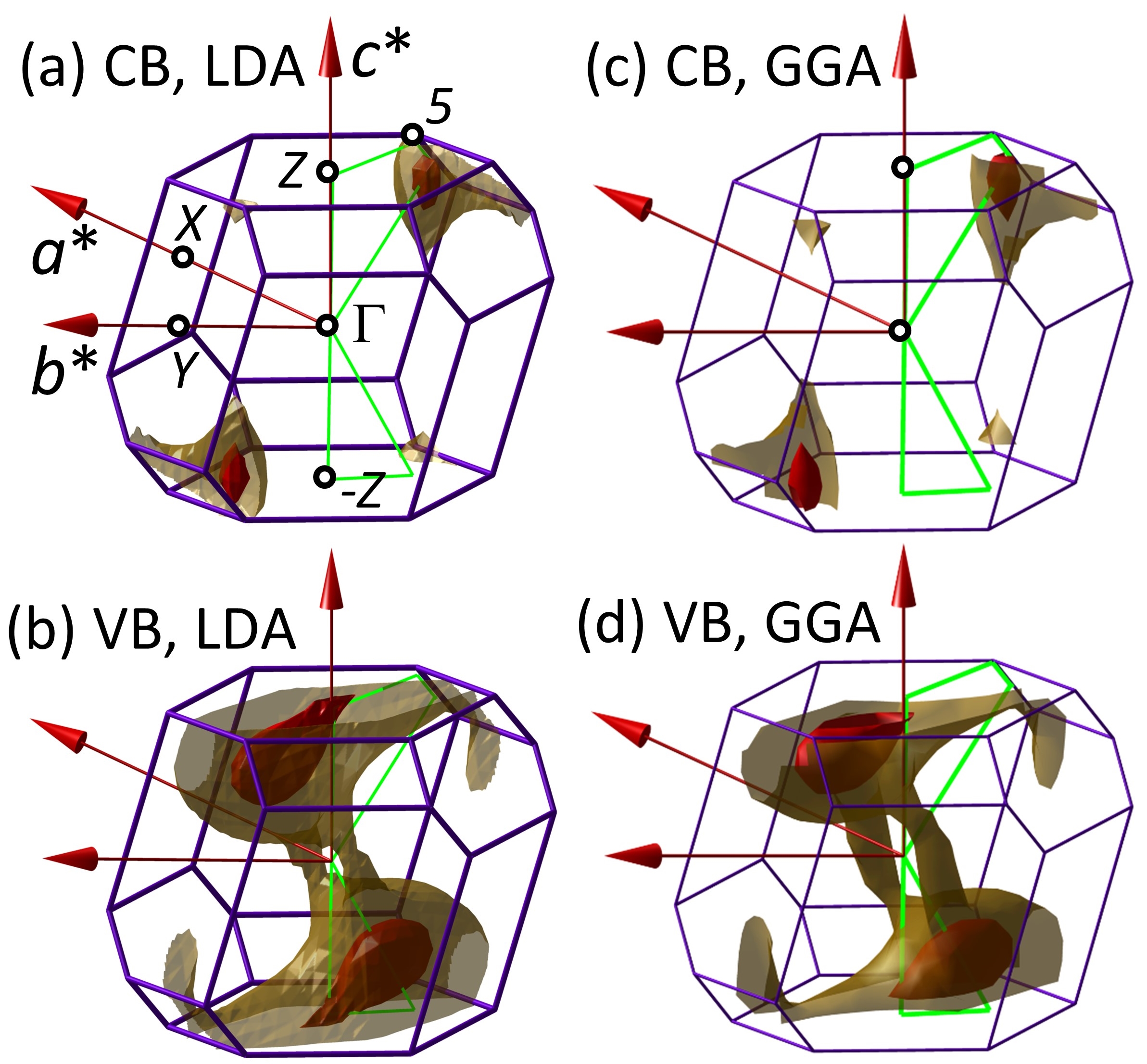}
\caption{ Constant energy surfaces in the Brillouin zone of ReS\textsubscript{2}. Calculations using LDA: (a) contours in the conduction band (CB) at energies of 70 meV (yellow) and 20 meV (red) above the CB minimum and (b)  contours in the valence band (VB) at energies of 310 meV (yellow) and 70 meV (red) below the VB maximum. Calculations using GGA: (c) conduction band at energies of 70 meV (yellow) and 20 meV (red) above the CB minimum and (d) valence band at energies of 220 meV (yellow) and 70 meV (red) below the VB maximum. The red arrows indicate the reciprocal lattice vectors; the vertical arrow corresponds to the \textit{\textit{c*}} axis which is normal to the layer planes and the sample surface. The green line indicates the path in the BZ used below in the LDA calculations of the band dispersion. The special points $X$, $Y$, $Z$, $-Z$ and \textit{5} [26-27] lying on the BZ boundary are indicated by circles.  }
\label{fig:figure2}
\end{figure}

\noindent The results of Figure 2 suggest that the lowest-energy inter-band transition is indirect in the bulk material, since the VBM and CBM do not coincide. This question has been actively discussed recently on the basis of optical, ARPES and transport measurements, and is not yet resolved. Briefly, for bulk ReS\textsubscript{2}, electron energy loss spectroscopy (EELS) results gave a room-temperature direct gap of 1.42 eV [24] whilst optical absorption-edge measurements suggested indirect gaps of 1.35 and 1.38 eV (for polarisations parallel and perpendicular respectively to the Re chains) [28] and electroluminescence data suggested an indirect gap of 1.41 eV [29]. Recently, a combined photoacoustic and modulated reflectance spectroscopy study concluded that both ReS\textsubscript{2} and ReSe\textsubscript{2} have indirect bandgaps of 1.37 and 1.18 eV respectively and higher-lying direct gaps of  1.55 and 1.31 eV [30]. Only one ARPES study has reported observation of the CB states via rubidium doping, and found at 10-20 K a direct gap located at the BZ boundary along the \textit{\textit{c*}} direction though with a substantially lower magnitude, of around 1.2 eV [13], assumed to be due to bandgap renormalisation at the high doping levels required. This special point in the BZ, labelled $Z$ [13, 20] (as in Fig. 2)  or $A$ [10], is the reported location of the VBM in all ARPES studies of ReS\textsubscript{2} to date, and this agrees reasonably well with the data of Fig. 2(b) and (d), though the figure demonstrates the flatness of this maximum so that, in our calculations, the VBM is actually displaced away from $Z$ to the centres of the lobes (red) which are constant energy surfaces 70 meV below the VBM; the precise location of the VBM is therefore challenging to determine via DFT, because discrepancies in energy of this magnitude easily arise from different choices of exchange-correlation functional and use of scalar \textit{versus} fully-relativistic pseudopotentials. To demonstrate the extent to which our choice of pseudopotential affects the conclusions, we show calculations using the LDA in Fig. 2 (a) and (b) and using the GGA in Fig. 2 (c) and (d). All the essential features of the constant energy surfaces are reproduced though there are minor differences in the exact energies and positions in the BZ of the band extrema; differences in energy between LDA and GGA global indirect band gaps are of the order of 100 to 200 meV. \newline

\noindent We note that our calculations are not in very good agreement with two other DFT calculations for bulk or many-layer ReS\textsubscript{2} which report a (T=0 K) direct gap of 1.35 eV at the $\bar{\Gamma}$ point [7,22], though we do find, in agreement with [13] that, along the \textit{\textit{c*}} direction, the VB has a \textit{local} maximum at the $Z$ point.This is most clearly seen in Fig. 2(b) and was reported earlier, where the direct gap at Z was found to be 1.525 eV [20]. It is also generally found that the VB at the true ${\Gamma}$ point shows a local minimum, giving an ‘$M$’-shaped dispersion, as shown by the bifurcation of the constant energy surface in Figure 2(b) and (d) (yellow). \newline

\noindent It is clear from Fig. 2(a) and (c) that the CB minimum in this level of approximation does not appear at $Z$ but is indeed located near the plane containing $Z$ and is displaced towards the 3D special point 5. This is in qualitative agreement with the calculations of [10] (their Figure 4d) and is potentially compatible with the ARPES results[13].  

\subsection*{Rhenium selenide}
\begin{figure}[t]
\centering
\includegraphics[width=0.7\textwidth]{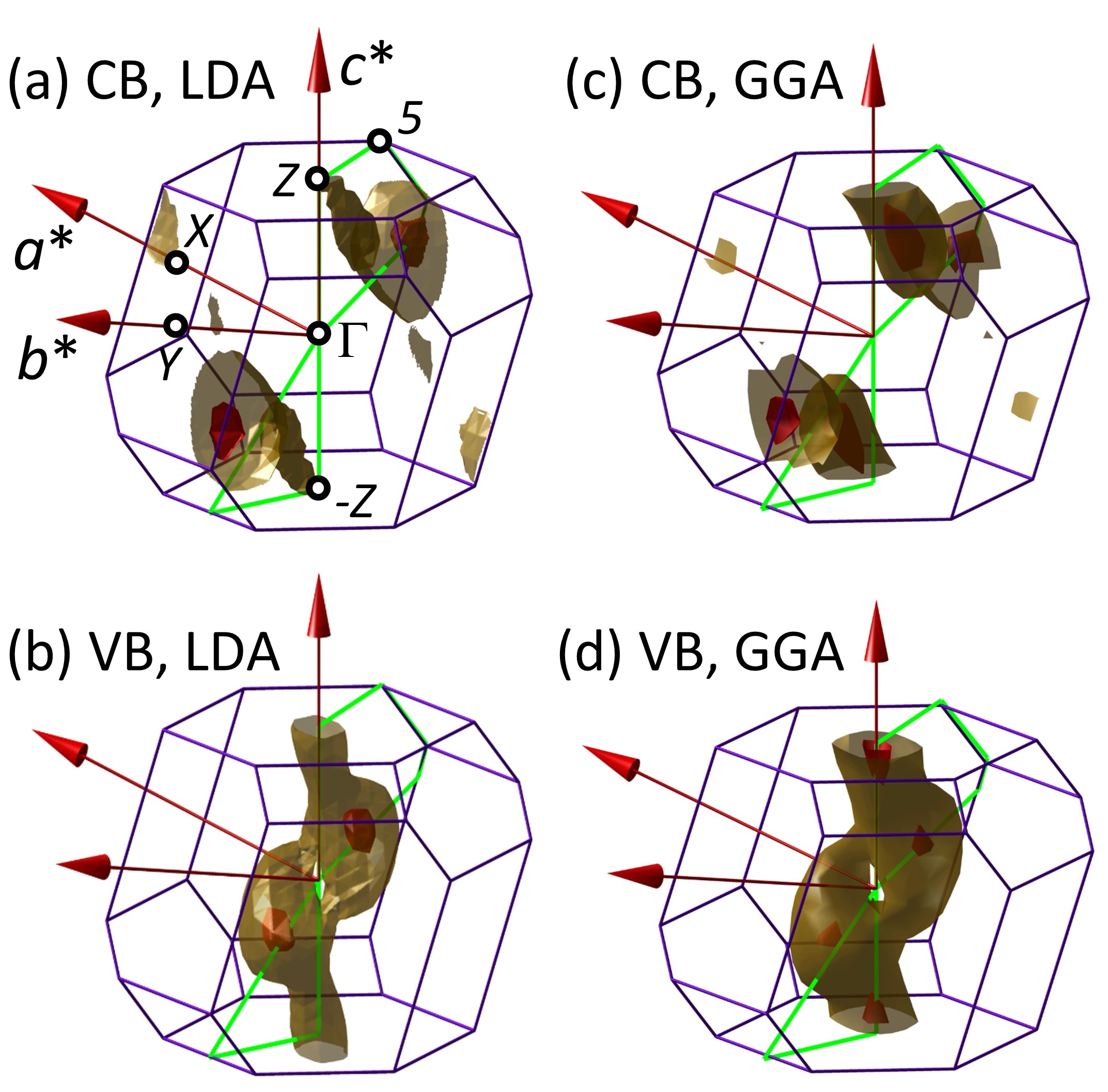}
\caption{ Constant energy surfaces in the Brillouin zone of ReSe\textsubscript{2}. Calculations using LDA: (a) contours in the conduction band (CB) at energies of 85 meV (yellow) and 20 meV (red) above the CB minimum and (b)  contours in the valence band (VB) at energies of 80 meV (yellow) and 15 meV (red) below the VB maximum. Calculations using GGA: (c) conduction band at energies of 70 meV (yellow) and 20 meV (red) above the CB minimum and (d) valence band at energies of 85 meV (yellow) and 20 meV (red) below the VB maximum. The red arrows indicate the reciprocal lattice vectors; the vertical arrow corresponds to the \textit{c*} axis which is normal to the layer planes and the sample surface. The green line again shows the path in the BZ used in the LDA calculation of the band dispersion. }
\label{fig:figure3}
\end{figure}
\noindent Fig. 3 shows the equivalent constant energy surfaces for ReSe\textsubscript{2} and, as might be expected from the chemical similarity between S and Se, the overall structure is similar to that of ReS\textsubscript{2}. Focusing first on the results obtained via the LDA, Fig. 3 (a) and (b), (i) we find the conduction band minimum again lies within the volume of the BZ (rather than at the surface or at $\Gamma$) and is closer to the plane containing $Z$ than to that containing $\Gamma$, and (ii) the valence band maxima are offset either side of the $\Gamma$ point. We have discussed ARPES data for ReSe\textsubscript{2} in detail elsewhere and reported already the form of one constant energy surface (yellow) for its VB [12] but we show this again in Fig. 3 (b) for comparison with the CB structure. We have also expanded Fig. 3 (b) to include a second constant energy surface (red) closer to the calculated VBM. Just as for ReS\textsubscript{2}, we find that the three-dimensional $\Gamma$ point is a local minimum in energy with a bifurcation of the constant energy surfaces around it. This is borne out by the experimental ARPES data [12] which shows that the VBM is displaced in the layer plane away from the projection of $\Gamma$ and agrees also with other DFT  calculations (Fig. S1 of the Supplementary material of ref. [23]). Turning to our results obtained in the GGA, we find that the same remarks apply, with one exception; in the GGA, two VB maxima appear, one in the same location as for the LDA results and one very close to $Z$. This does not, however, affect the conclusion that ReSe\textsubscript{2} is indirect, since the CB minimum is still not co-located with the VBM; we find global indirect gaps in LDA, GGA respectively of 0.87 eV, 0.99 eV and a direct gap at $Z$ of 0.97 eV, 1.00 eV. One other recent calculation gives an indirect gap of 0.92 eV for bulk ReSe\textsubscript{2} in the LDA with a direct quasiparticle gap at $Z$ of 1.49 in the LDA+GdW approximation [27]; a second calculation in the GGA+GW approximation gives a direct gap at $Z$ of 1.38 eV [26]. The recent photoacoustic spectroscopy study gives an indirect gap of 1.18 eV and direct gap of 1.31 eV [30]. However, both computational studies only report band dispersions along specific paths in the BZ and it would be interesting in future to apply these more sophisticated models to the whole BZ.  \newline
 
\noindent The findings of the present calculations are summarised in Figure 4, which shows the band energies for ReS\textsubscript{2} and ReSe\textsubscript{2} taking a path in the Brillouin zone passing through the key points. Again, calculations for both LDA and GGA are shown. To illustrate this, the path used for the LDA calculations is shown (green lines) in Figs. 2 and 3 and is as follows: starting from $\Gamma$, it follows a straight line through the CB minimum (CBM) and continues along this line to the BZ boundary (we label this point CBM'). It then turns to run along the BZ surface up to point 5 in the plane containing the $Z$ point, crosses that plane to $Z$ and runs back down to $\Gamma$. The energies of both CB and VB states are plotted for this path. This accounts for the left hand halves of Fig. 4(a-d); in the right hand halves, we then follow the same procedure, from $\Gamma$ to the VB maximum, on to the BZ boundary (VBM') and then across the BZ surface to $Z$ and finally back to $\Gamma$. This shows clearly that the VBM and CBM do not coincide so that, considering first ReS\textsubscript{2}, Fig. 2 (a) and (b), the material is formally an indirect semiconductor, but that there is a gap at $Z$ which is formed between a local VB maximum and a very flat CB minimum and is therefore a slightly larger direct gap. In the region of the VBM, it appears that there is also a close-lying local minimum in the CB, providing another nearly direct gap. It is hardly surprising, therefore, that measurements of optical reflection or absorption have difficulty in identifying the nature of the band gap. Similar conclusions follow from the analogous data for ReSe\textsubscript{2} shown in Fig. 4 (c) and (d).  

\begin{figure}[t]
\centering
\includegraphics[width=0.9\textwidth]{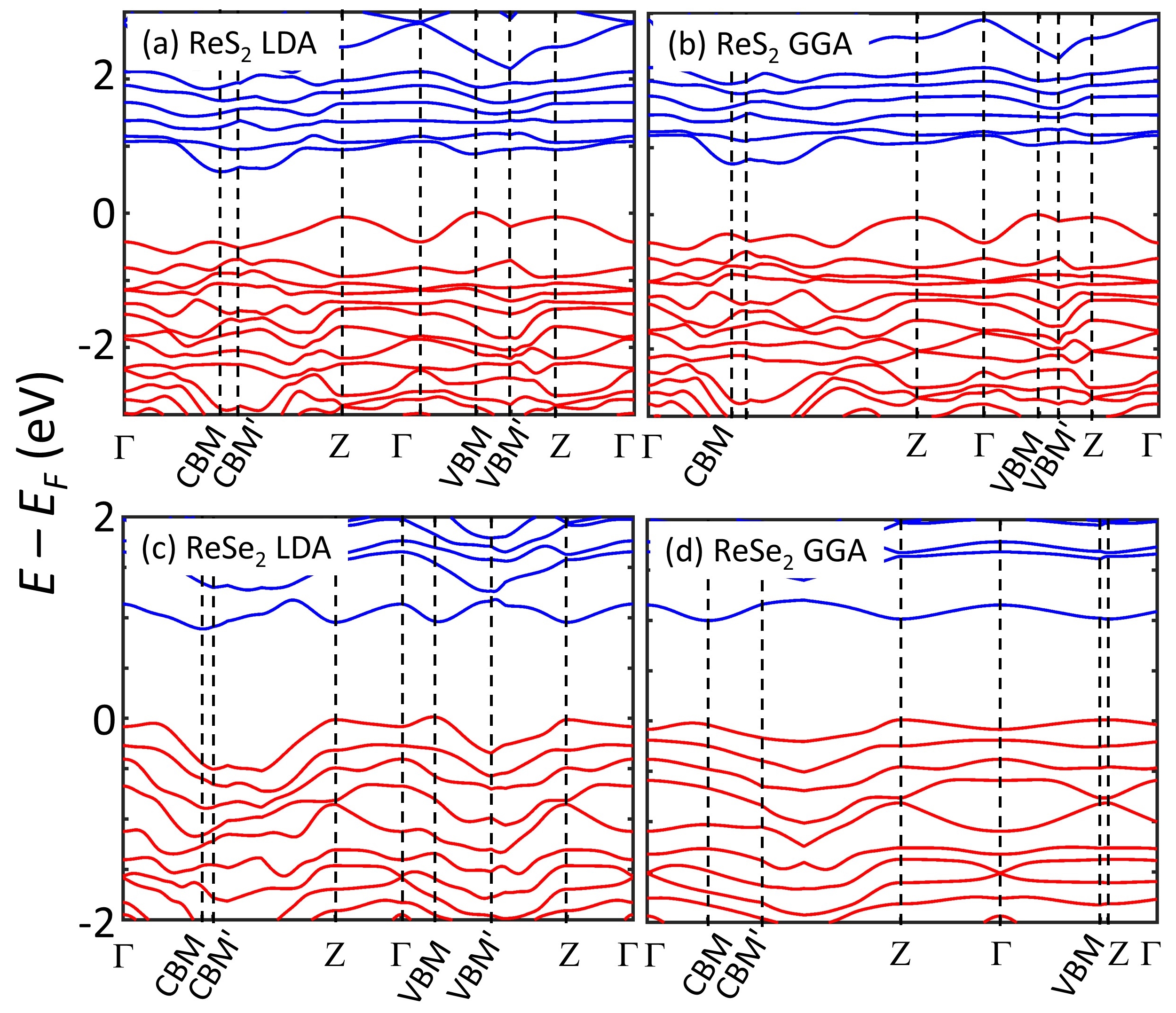}
\caption{Conduction and valence band edges for (a) and (b) ReS\textsubscript{2} and (c) and (d) ReSe\textsubscript{2} for a path around the Brillouin zone passing through the $\Gamma$ and $Z$ points and the band extrema (CBM and VBM). The other labels (CBM', VBM') are explained in the text. The paths through the Brillouin zone used for the LDA calculation are indicated by the green lines in Figures 2 and 3.  Data of (a), (b) were calculated using LDA exchange-correlation functionals and GGA functionals were used for (c), (d).}
\label{fig:figure4}
\end{figure}

\section*{CONCLUSIONS}
\noindent We have presented a detailed comparison of the band structures of bulk ReS\textsubscript{2} and ReSe\textsubscript{2} calculated using LDA and GGA DFT with fully relativistic pseudopotentials and use of the PAW method. We show that only a calculation of the band structure over the whole three-dimensional Brillouin zone volume can identify correctly the locations of the band extrema for a given level of computational accuracy. Our results show clearly that the task of classifying the interband optical transitions in ReS\textsubscript{2} \textit{via} DFT calculations is very demanding since direct and indirect transitions lie close in energy and within the range of DFT results arising from different choices of pseudopotential. In the case of ReSe\textsubscript{2}, computational and experimental results are in better agreement though, it should be noted, there is less experimental data available; the current consensus is that bulk ReSe\textsubscript{2} is an indirect semiconductor.

\begin{acknowledgments}
\noindent This work was supported by the Centre for Graphene Science of the Universities of Bath and Exeter and by the Engineering and Physical Sciences Research Council EPSRC (UK) under Grants No.~EP/G036101, No.~EP/M022188, and No.~EP/P004830; S.~M.~G.~and L.~S.~H.~are supported by the Bath-Bristol Centre for Doctoral Training in Condensed Matter Physics, Grant No.~EP/L015544. Associated experimental studies were supported by the award of beam time at the DIAMOND (IO5) and SOLEIL (ANTARES) synchrotron beam lines and by EPSRC Grant No.~EP/P004830/1. Computational work was performed on the University of Bath’s High Performance Computing Facility. Data created during this research are freely available from the University of Bath data archive at \href{http://dx.doi.org/10.15125/BATH-00331}{DOI:10.15125/BATH-00331}, \href{http://dx.doi.org/10.15125/BATH-00332}{DOI:10.15125/BATH-00332}.
\end{acknowledgments}

\section*{REFERENCES}
\noindent[1] K. S. Novoselov, D. Jiang, F. Schedin, T. J. Booth, V. V. Khotkevich, S. V. Morozov and A. K. Geim, PNAS 102, 10451 (2005) \newline
[2]	J. A. Wilson and A. D. Yoffe, Adv. Phys. 18, 193 (1969) \newline
[3]	Jin-Wu Jiang, Frontiers of Physics 10, 287 (2015) \newline
[4]	S. Manzeli, D. Ovchinnikov, D. Pasquier, O. V. Yazyev and A. Kis, Nat. Reviews Materials 2, 17033 (2017) \newline
[5]	M. Rahman, K. Davey and S‐Z. Qiao, Adv. Funct. Mater. 27, 1606129 (2017) \newline
[6]	H. J. Lamfers, A. Meetsma, G. A. Wiegers and J. L. deBoer, J. Alloys Compd. 241, 34 (1996) \newline
[7]	S. Tongay, H. Sahin, C. Ko, A. Luce, W. Fan, K. Liu, J. Zhou, Y. S. Huang, C. H. Ho, J. Y. Yan, D. F. Ogletree, S. Aloni, J. Ji, S. S. Li, J. B. Li, F. M. Peeters and J. Q. Wu, Nat. Commun. 5, 3252 (2014) \newline
[8]	D. Wolverson, S. Crampin, A. S. Kazemi, A. Ilie and S. J. Bending, ACS Nano 8, 11154 (2014) \newline
[9]	E. Canadell, A. LeBeuze, M. Abdelaziz El Khalifa, R. Chevrel and Myung Hwan Whangbo, J. Amer. Chem. Soc. 111, 3778 (1989) \newline
[10] M. Gehlmann, I. Aguilera, G. Bihlmayer, S. Nemšák, P. Nagler, P. Gospodarič, G. Zamborlini, M. Eschbach, V. Feyer, F. Kronast, E. Młyńczak, T. Korn, L. Plucinski, C. Schüller, S. Blügel and C. M. Schneider, Nano Lett. 17, 5187 (2017) \newline
[11] J. L. Webb, L. S. Hart, D. Wolverson, C. Y. Chen, J. Avila and M. C. Asensio, Phys. Rev. B 96, (2017) \newline
[12] L. S. Hart, J. L. Webb, S. Dale, S. J. Bending, M. Mucha-Kruczynski, D. Wolverson, C. Y. Chen, J. Avila and M. C. Asensio, Sci Rep 7, 5145 (2017) \newline
[13] D Biswas, Alex M Ganose, R Yano, JM Riley, L Bawden, OJ Clark, J Feng, L Collins-Mcintyre, MT Sajjad and W Meevasana, Phys. Rev. B 96, 085205 (2017) \newline
[14] P. Giannozzi, S. Baroni, N. Bonini, M. Calandra, R. Car, C. Cavazzoni, D. Ceresoli, G. L. Chiarotti, M. Cococcioni, I. Dabo, A. Dal Corso, S. de Gironcoli, S. Fabris, G. Fratesi, R. Gebauer, U. Gerstmann, C. Gougoussis, A. Kokalj, M. Lazzeri, L. Martin-Samos, N. Marzari, F. Mauri, R. Mazzarello, S. Paolini, A. Pasquarello, L. Paulatto, C. Sbraccia, S. Scandolo, G. Sclauzero, A. P. Seitsonen, A. Smogunov, P. Umari and R. M. Wentzcovitch, J Phys-Condens Mat 21, 395502 (2009) \newline
[15] G. Kresse and D. Joubert, Phys. Rev. B 59, 1758 (1999) \newline
[16] A. Dal Corso, Comput. Mater. Sci. 95, 337 (2014) \newline
[17] J. P. Perdew and Alex Zunger, Phys. Rev. B 23, 5048 (1981) \newline
[18] J. P. Perdew, K. Burke and M. Ernzerhof, Phys. Rev. Lett. 77, 3865 (1996) \newline
[19] H. J. Monkhorst and J. D. Pack, Phys. Rev. B 13, 5188 (1976) \newline
[20] J. L. Webb, L. S. Hart, D. Wolverson, C. Chen, J. Avila and M. C. Asensio, Phys. Rev. B 96, 115205 (2017) \newline
[21] C. H. Ho, Y. S. Huang, P. C. Liao and K. K. Tiong, J. Phys. Chem. Solids 60, 1797 (1999) \newline
[22] E. Liu, Y. Fu, Y. Wang, Y. Feng, H. Liu, X. Wan, W. Zhou, B. Wang, L. Shao and C.-H. Ho, Nat. Commun. 6, 6991 (2015) \newline
[23] H. Zhao, J. Wu, H. Zhong, Q. Guo, X. Wang, F. Xia, L. Yang, P. Tan and H. Wang, Nano Research 8, 3651 (2015) \newline
[24] K Dileep, R Sahu, Sumanta Sarkar, Sebastian C Peter and Ranjan Datta, J. Appl. Phys. 119, 114309 (2016) \newline
[25] W. Wen, Y. Zhu, X. Liu, H. P. Hsu, Z. Fei, Y. Chen, X. Wang, M. Zhang, K. H. Lin, F. S. Huang, Y. P. Wang, Y. S. Huang, C. H. Ho, P. H. Tan, C. Jin and L. Xie, Small 13, 1603788 (2017) \newline
[26] J. P. Echeverry and I. C. Gerber, Phys. Rev. B 97, 075123 (2018) \newline
[27] A. Arora, J. Noky, M. Drüppel, B. Jariwala, T. Deilmann, R. Schneider, R. Schmidt, O. Del Pozo-Zamudio, T. Stiehm, A. Bhattacharya, P. Krüger, S. Michaelis de Vasconcellos, M. Rohlfing and R. Bratschitsch, Nano Lett. 17, 3202 (2017) \newline
[28] C. H. Ho, Y. S. Huang, K. K. Tiong and P. C. Liao, Phys. Rev. B 58, 16130 (1998) \newline
[29] I. Gutierrez-Lezama, B. A. Reddy, N. Ubrig and A. F. Morpurgo, 2D Materials 3, (2016) \newline
[30] S. J. Zelewski and R. Kudrawiec, Sci Rep 7, 15365 (2017) \newline
\newline

\end{document}